\documentstyle[preprint,aps]{revtex}
\tighten
\begin{document}
\draft
\hyphenation{Rijken}
\hyphenation{Nijmegen}

\title{Comparison of potential models with the
       \protect\bbox{\lowercase{np}} scattering data below 350 MeV}

\author{Vincent Stoks}
\address{School of Physical Sciences, The Flinders University of
         South Australia, \\ Bedford Park, South Australia 5042, Australia}
\author{J.J.\ de Swart}
\address{Institute for Theoretical Physics, University of Nijmegen,
         Nijmegen, The Netherlands}

\date{}
\maketitle

\begin{abstract}
As a follow-up on our earlier paper we calculate the $\chi^{2}$ of
various $N\!N$ potential models with respect to the $np$ scattering data.
We find that only the most recent potential models give a reasonably
good description of these data. Almost none of the potentials is truly
an $N\!N$ potential in the sense that it gives a good description of
both $pp$ and $np$ scattering simultaneously.
\end{abstract}

In a recent paper~\cite{St93a}, we investigated the quality of a number
of $N\!N$ potential models with respect to the $pp$ scattering data
below 350 MeV. By comparing to the $pp$ data we tested
the isovector ($I=1$) partial waves of these models.
In this Brief Report we will confront the models with the $np$ data
so that we can also test the isoscalar ($I=0$) partial waves.

In our previous comparison~\cite{St93a}, we restricted ourselves to a
comparison with respect to the $pp$ data only, because of two reasons.
Firstly, at the time of our comparison we were still in the process of
analyzing the $np$ data, so we did not have a complete and consistent
$np$ database. Secondly, our initial goal for comparing $N\!N$ models
was to bring attention to the fact that these models are often not
$N\!N$ models at all, but rather $pp$ or $np$ models; i.e., they were
only fitted to one type ($pp$ or $np$) of data.
Especially in the last decade, it has been assumed by many people
that it is sufficient to fit the parameters of a potential model only
to the $np$ data. Simply adding the proper electromagnetic interaction
is then assumed to provide the corresponding $pp$ version of that
potential, so that one in fact ends up with an $N\!N$ potential.
(This is much easier than fitting to the $pp$ data directly, because
in fitting only the $np$ data one does not have to worry about all
kinds of electromagnetic corrections when calculating the observables.)
For example, the full Bonn potential~\cite{Mac87} was originally
presented as an $N\!N$ potential, although its parameters were in fact
fitted only to the $np$ data. In a later publication~\cite{Hai89}, the
model had to be adjusted to make it also applicable to $pp$ scattering.
We wanted to show explicitly that in order to construct an $N\!N$
potential model it is generally not good enough to only fit the $np$
data.

One often argues incorrectly that the reason for this failure is only
artificial, where the argument goes as follows. There are some very
accurate $pp$ differential cross sections~\cite{Tho78} and $np$ total
cross sections~\cite{Hou71} at low energies, which put very tight
constraints on the $pp$ and $np$ $^{1}S_{0}$ scattering lengths.
It is well known that the nuclear $pp$ and $np$ scattering lengths
are different and that this difference cannot be explained by mass
differences alone. It is then obvious that the choice of fitting a
potential to either the $pp$ or $np$ scattering length will result in
a totally incorrect $np$ or $pp$ scattering length, respectively, for
that model. Very high $\chi^{2}$ values on the $pp$ data for
models which are fitted to the $np$ scattering length will then be
the result.

A way to avoid this problem is to exclude these low-energy data and
only include data above 5 MeV, say. Fitting to either the $pp$ or $np$
scattering length causes only minor differences in the quality of the
description of the data at energies above 5 MeV. However, in doing this
we found~\cite{St93a} that most of the potential models still gave a
very poor description of the $pp$ data. This indicates that the poor
quality of these models is not only due to an incorrect scattering
length, but has other sources as well.

Another argument that a high $\chi^{2}$ on the $pp$ data is only
artificial is that one can slightly adjust one of the parameters of
the potential model to ensure that it fits the $pp$ $^{1}S_{0}$
scattering length. Because the scattering length is extremely sensitive
to small changes in the potential, one expects that such an adjustment
will be minimal and will hardly affect the other partial waves.
This modification is then believed to improve the description of the
low-energy data without affecting the description of the other data,
resulting in an overall good description of the $pp$ data.
We investigated this~\cite{St93a} extensively using the Arg84
potential~\cite{Wir84}. Instead of modifying one of the parameters,
we completely replaced the $pp$ $^{1}S_{0}$ phase shift of the Arg84
potential by the $pp$ $^{1}S_{0}$ phase shift of the Nijmegen
partial-wave analysis~\cite{Ber90}. This corresponds to having an
almost ``perfect'' $^{1}S_{0}$ phase shift; something which is very
unlikely to be achievable by simply adjusting one of the potential
parameters. Although this gives a considerable improvement, the
resulting $\chi^{2}/N_{\rm data}$ is still rather high. The same trend
is found when we apply this procedure to the coordinate-space version
of the Bonn potential~\cite{Mac87} or to the coordinate-space Bonn A
and Bonn B potentials~\cite{Mac89}.
This indicates that the poor quality of these potential models with
respect to the $pp$ data is not only due to the fact that they have
an incorrect $^{1}S_{0}$ phase shift. In fact, because of the mass
difference between neutral and charged pions the $pp$ and $np$
isovector phase shifts in the other partial waves are significantly
different as well (see Table I of Ref.~\cite{St93b}).
Therefore, the problem usually cannot be fixed by simply adjusting
one of the parameters of the model.

Now that the Nijmegen partial-wave analysis of the $np$ data below
350 MeV is finished~\cite{St93b}, we have a complete and consistent
$np$ database, and we can complete our comparison of $N\!N$ potential
models by confronting them with the $np$ data.
This allows us to investigate whether $np$ potential models which
give a very poor description of the $pp$ data can in fact survive a
quality test with respect to the $np$ data. As we will see, some of
them do. This once more strongly supports our statement that
``a good fit to the $np$ data does not automatically guarantee a
good fit to the $pp$ data''~\cite{St93a}.

In this Brief Report we will confront a number of potential models
with the $np$ scattering data. This means that we have to calculate the
phase shifts of the lower partial waves at all the energies at which
experimental data are available, construct the scattering amplitude,
and calculate the observables. Since we are only going up to
$T_{\rm lab}=350$ MeV, it is sufficient to calculate the phase shifts
due to the nuclear potential up to total angular momentum $J=6$.
For the higher partial waves we can take the phase shifts as given
by one-pion exchange.
The $np$ database we use is given in Ref.~\cite{St93b}, with the
difference that here we will leave out the low-energy data below
$T_{\rm lab}=5$ MeV as discussed earlier. The database now consists
of 2458 data between 5 and 350 MeV.

The potential models we will consider are mainly the same as those
of our $pp$ comparison~\cite{St93a}, namely: HJ62~\cite{Ham62},
Reid68~\cite{Rei68}, TRS75~\cite{Tou75}, Paris80~\cite{Lac80},
Urb81~\cite{Lag81}, Arg84~\cite{Wir84}, BonnR~\cite{Mac87},
Bonn89~\cite{Mac87,Hai89}, and Nijm93~\cite{Sto94}.
The soft-core Nijmegen potential is here referred to as Nijm93.
This is an updated version of the Nijm78 potential~\cite{Nag78}.

We made a change in notation with respect to the Bonn potentials,
in that the coordinate-space Bonn potential is now denoted by BonnR,
and Bonn89 refers to the full Bonn potential.
Because we do not have the computer codes to calculate the phase
shifts of the full Bonn potential ourselves, these phase shifts were
obtained from the computer software SAID~\cite{said}. We take the
phase shifts in steps of 1 MeV as generated by the SAID program,
and then use linear interpolation to obtain the phase shifts at the
experimental energies. Incidentally, this provides another reason for
only using the $np$ data above 5 MeV: The phase shifts below 10 MeV in
the SAID program are represented by an effective-range parametrization
which, especially for the $S$ waves, is not good enough to represent
the low-energy phase shifts to a high accuracy. We checked for the
other potentials that starting at 5 MeV rather than at 10 MeV does not
make any difference for the conclusions we will draw below.

As a check that the way how we treat the full Bonn potential (Bonn89)
is justified, we applied the same procedure to the Nijmegen
potential: We took the phase shifts of the Nijm93 potential
in steps of 1 MeV as generated by the SAID program, interpolated to
get the phase shifts at the experimental energies, calculated the total
$\chi^{2}$ with respect to the $np$ data, and compared to the result
we obtain when we calculate the Nijm93 phase shifts directly using our
own computer code for the Nijm93 potential.
The difference in $\chi^{2}$ is less than 0.3. This strongly supports
our belief that for the present purpose we can use the phase shifts
of the Bonn89 potential as stored in the SAID program.

The results for the nine potential models are presented in
Table~\ref{chinp}.
Similarly to our comparison of potential models with respect to the
$pp$ scattering data~\cite{St93a}, only a few of the models give a
reasonable description of the $np$ scattering data. The best ones
are the Arg84 and Nijm93 models with $\chi^{2}/N_{\rm data}\approx2$,
followed by the Urb81 and Bonn89 models with a still reasonable
$\chi^{2}/N_{\rm data}\approx3$.
To get some insight in where the high $\chi^{2}$ for the other models
comes from, we divide the total $\chi^{2}$ in a set of sub-$\chi^{2}$
for each different type of observable, as listed in Table~\ref{chinp}.

The relatively high $\chi^{2}(\sigma_{\rm tot})$ for most models
is mainly due to one group of total cross sections~\cite{Lis82},
consisting of 70 data from 39 to 350 MeV. The statistical error on
these data is rather small and the energy range is very large. So if
the energy dependence (the shape) for the total cross sections as given
by a potential model is different from what it is implied to be according
to the experiment, the $\chi^{2}$ rapidly increases. Indeed, the other
total cross-section experiments cover much smaller energy ranges and
the description of these data is generally much better with
$\chi^{2}/N_{\rm data}$ ranging from 1.1 for the Arg84 potential to
3.7 for the BonnR potential. However, there still remains the fact
that this group of 70 data can be described very well in the Nijmegen
partial-wave analysis~\cite{St93b} with $\chi^{2}=56.8$.

High $\chi^{2}(\sigma(\theta))$ and $\chi^{2}(A_{yy},A_{zz})$ indicate
an incorrect $^{1}P_{1}$ phase shift and $\epsilon_{1}$ mixing parameter.
For the HJ62, Reid68, and Urb81 models the $^{1}P_{1}$ phase shift
becomes too negative at high energies, whereas for the BonnR model
it does not become negative enough.
At $T_{\rm lab}=300$ MeV the Nijmegen partial-wave analysis~\cite{St93b}
gives $\delta(^{1}P_{1})=-27.58(22)^{\circ}$.
Similarly, comparing to~\cite{St93b} $\epsilon_{1}=4.03(17)^{\circ}$,
the $\epsilon_{1}$ mixing parameter is too high in the HJ62, Reid68, and
TRS75 models, whereas it is too small in the Bonn models.

As an example of the differences between the phase shifts of the various
potential models, we show in Fig.~\ref{figphs} the $\epsilon_{1}$
mixing parameter and the $^{3}D_{2}$ phase shift. The solid curves
are the values from the multi-energy Nijmegen partial-wave
analysis~\cite{St93b}. The shaded band represents the statistical error.
Compared to the Nijmegen analysis, the Paris80 potential gives values
which are too high, whereas the Bonn89 potential gives values which
are too low. This is reflected in the relatively high $\chi^{2}$ on
the differential cross sections and spin-correlation paremeters for
these models.

In Table~\ref{chinp} we also give the $\chi^{2}/N_{\rm data}$ for
the 5--350 MeV range and 0--350 MeV energy ranges.
The latter is included to clearly demonstrate the enormous changes that
can occur when we include the low-energy data. Obviously, including
these low-energy data in a potential comparison can be misleading in
that the large total $\chi^{2}$ obscures the fact that a certain
potential model can in fact describe the scattering data at energies
above 5 MeV, say, reasonably well.

To summarize, we find that only the Arg84 and Nijm93 potentials give a
good description of the $np$ data ($\chi^{2}/N_{\rm data}\approx2$),
while the Urb81 and Bonn89 potentials can be qualified as reasonable
($\chi^{2}/N_{\rm data}\approx3$).
When we include the very accurate low-energy total cross sections,
only the Nijm93 potential still has a good $\chi^{2}/N_{\rm data}$.

Before we can make statements about the general quality of $N\!N$
potential models, we have to recapitulate the quality of these models
with respect to the $pp$ scattering data. Rather than referring to
our previous publication~\cite{St93a}, we here present the results for
the $pp$ data using the same procedure we use for the $np$ data.
So also for the $pp$ data, we here give $\chi^{2}/N_{\rm data}$ results
from a direct comparison to the $pp$ data, rather than using the
Nijmegen representation of the $\chi^{2}$ hypersurface of the $pp$ data.
The results are given in Table~\ref{chipp}.
Note once more the enormous rise for some of the models (Arg84 and BonnR)
when we include the low-energy (0--5 MeV) data.
The quality with respect to the $pp$ data for each model individually
has already been discussed in our previous publication~\cite{St93a}.

Comparing Tables~\ref{chinp} and \ref{chipp}, we conclude that most
potential models are not $N\!N$ models at all, but give only a reasonable
description of either the $pp$ or the $np$ scattering data (or not even
that). The only exception is the Nijm93 potential. The situation for
the Bonn89 potential (except at very low energies) is probably also still
not unsatisfactory, although the description of the $np$ data is
considerably worse (but not too bad) than the description of the $pp$
data. The TRS75 potential also shows a quality which is consistent for
both $pp$ and $np$ scattering, although $\chi^{2}/N_{\rm data}=3.5$ on
all data is not too good.
All other potentials can be classified as either $pp$ potentials
(Reid68, Paris80), $np$ potentials (Urb81, Arg84), or they do not
give a satisfactory description of either $pp$ or $np$ scattering
(HJ62, BonnR). Two other coordinate-space versions of the Bonn
potentials, Bonn A and Bonn B~\cite{Mac89}, also belong to the last
category. For the 5--350 MeV energy range, $\chi^{2}/N_{\rm data}$ for
these two models is: for Bonn A, 9.4 on $pp$ and 8.3 on $np$, and for
Bonn B, 8.5 on $pp$ and 8.9 on $np$. It is surprising to see the
enormous difference in quality of the full Bonn potential (Bonn89) on the
one hand and all the Bonn coordinate-space versions on the other hand.
The difference cannot be explained~\cite{Mac87,Mac89} by claiming that
coordinate-space potentials are necessarily of inferior quality.
The quality of the Nijmegen potential, which is a coordinate-space
potential (but which also has an exactly equivalent momentum-space
version), clearly contradicts this.

The results from Tables~\ref{chinp} and \ref{chipp} are rather
disappointing, considering the fact that all models were originally
presented as being $N\!N$ potentials.
This demonstrates once again our point that in general one has to be
careful when using these potential models in other calculations, like
in few-nucleon scattering and bound-state calculations, $pp$
bremsstrahlung, or nuclear matter calculations.
In most cases these models cannot even describe both $pp$ and $np$
scattering with the same, satisfactory, quality.

We should mention that recently there have been constructed a number
of new $N\!N$ potentials which are truly $N\!N$ potentials in the sense
that they give an excellent description of both the $pp$ and $np$ data
simultaneously. These are two Nijmegen potentials Nijm~I and
Nijm~II~\cite{Sto94}, a regularized update of the old Reid68
potential~\cite{Sto94}, and an update of the old Arg84
potential~\cite{Wir94}. All four models have the almost optimal
$\chi^{2}/N_{\rm data}\approx1$ on both $pp$ and $np$ data.
Part of the success of these models is that they explicitly contain
the one-pion-exchange potential with the proper neutral- and
charged-pion masses. Another reason is of course that these models
were explicitly fitted to both $pp$ and $np$ data simultaneously
giving the proper constraints on the isovector partial waves.

The work of V.\ Stoks was supported by the Australian Research Council.
V.\ Stoks is also grateful to I.R.\ Afnan for discussions and comments
on the manuscript.

\begin{table}
\caption{$\chi^{2}$ on the $np$ scattering data between 5 and 350 MeV
         for various $N\!N$ potential models. A division is made showing
         the sub-$\chi^{2}$ on the total cross sections, the differential
         cross sections, the analyzing powers, the spin-correlation
         parameters, the depolarizations, and the rotation parameters.
         The lower part shows the $\chi^{2}/N_{\rm data}$ for the
         5--350 MeV and 0--350 MeV energy ranges.}
\begin{tabular}{c|r|rrrrrrrrr}
  & $N_{\rm data}$ & HJ62  & Reid68 & TRS75  &          Paris80
                   & Urb81 & Arg84  & BonnR  & Bonn89 & Nijm93  \\
  \tableline
 $\sigma_{\rm tot}$ &  225 &  2376 &  6599 &   389 &         1117
                           &   594 &   662 &  1624 &   895 &  554 \\
 $\sigma(\theta)$   & 1323 &  3379 &  5771 &  3273 &         2856
                           &  2726 &  2055 & 14908 &  3138 & 1554 \\
 $A_{y}$            &  738 &  2345 &  4217 &  3799 &         1275
                           &  1807 &  2019 &  3330 &  1791 & 1023 \\
 $A_{yy}, A_{zz}$   &   86 &   975 &  9456 &  1054 &         3899
                           &  1331 &   231 &  5822 &  1328 & 1499 \\
 $D_{t}$            &   43 &    56 &   133 &    87 &          108
                           &    81 &    51 &   296 &    90 &   52 \\
 $A_{t}, R_{t}$     &   43 &    60 &   127 &   176 &          202
                           &   145 &    82 &   348 &    89 &  100 \\
 all data           & 2458 &  9190 & 26303 &  8778 &         9457
                           &  6684 &  5100 & 26328 &  7330 & 4783 \\
 & & & & & & & & & & \\
  5--350 MeV & 2458 &  3.7 &  10.7 &   3.6 &          3.8
                    &  2.7 &   2.1 &  10.7 &   3.0 &  1.9 \\
  0--350 MeV & 2514 & 4100 &  3740 &  2925 &         3300
                    & 3400 &   3.3 &  10.5 &   423 &  1.9
\end{tabular}
\label{chinp}
\end{table}

\begin{table}
\caption{$\chi^{2}/N_{\rm data}$ on the $pp$ scattering data for the
         5--350 MeV and 0--350 MeV energy ranges.}
\begin{tabular}{c|r|rrrrrrrrr}
  & $N_{\rm data}$ & HJ62  & Reid68 & TRS75  &          Paris80
                   & Urb81 & Arg84  & BonnR  & Bonn89 & Nijm93  \\
  \tableline
  5--350 MeV & 1590 &  9.7 &   2.5 &   3.3 &          2.2
                    &  5.9 &   6.9 &  12.4 &   1.8 &  1.9 \\
  0--350 MeV & 1787 & 13.5 &   2.9 &   3.4 &          4.5
                    &  6.0 &  7615 &  1090 &  25.5 &  1.8
\end{tabular}
\label{chipp}
\end{table}

\begin{figure}
\caption{$\epsilon_{1}$ mixing parameter and $^{3}D_{2}$ phase shift
         of the Nijmegen partial-wave analysis with the statistical
         error (shaded band), the Paris potential (dotted line), the
         full Bonn potential (dash-dotted line), and the Nijmegen
         potential (dashed line).}
\label{figphs}
\end{figure}


\begin{references}
\bibitem{St93a} V.\ Stoks and J.J.\ de Swart,
         Phys.\ Rev.\ C {\bf 47}, 512 (1993).
\bibitem{Mac87} R.\ Machleidt, K.\ Holinde, and Ch.\ Elster,
         Phys.\ Rep.\ {\bf 149}, 1 (1987).
\bibitem{Hai89} J.\ Haidenbauer and K.\ Holinde,
         Phys.\ Rev.\ C {\bf 40}, 2465 (1989).
\bibitem{Tho78} Ch.\ Thomann, J.E.\ Benn, and S.\ M\"{u}nch,
         Nucl.\ Phys.\ {\bf A303}, 457 (1978).
\bibitem{Hou71} T.L.\ Houk, Phys.\ Rev.\ C {\bf 3}, 1886 (1971);
                W.\ Dilg, Phys.\ Rev.\ C {\bf 11}, 103 (1975);
                Y.\ Fujita, K.\ Kobayashi, T.\ Oosaki, and R.C.\ Block,
                Nucl.\ Phys.\ {\bf A258}, 1 (1976).
\bibitem{Wir84} R.B.\ Wiringa, R.A.\ Smith, and T.L. Ainsworth,
         Phys.\ Rev.\ C {\bf 29}, 1207 (1984).
\bibitem{Ber90} J.R.\ Bergervoet, P.C.\ van Campen, R.A.M.\ Klomp,
         J.-L.\ de Kok, T.A.\ Rijken, V.G.J.\ Stoks, and J.J.\ de Swart,
         Phys.\ Rev.\ C {\bf 41}, 1435 (1990).
\bibitem{Mac89} R.\ Machleidt, Adv.\ Nucl.\ Phys.\ {\bf 19}, 189 (1989).
\bibitem{St93b} V.G.J.\ Stoks, R.A.M.\ Klomp, M.C.M.\ Rentmeester,
         and J.J.\ de Swart,
         Phys.\ Rev.\ C {\bf 48}, 792 (1993).
\bibitem{Ham62} T.\ Hamada and I.D.\ Johnston, Nucl.\ Phys.\ {\bf 34},
         382 (1962); T.\ Hamada, Y.\ Nakamura, and R.\ Tamagaki,
         Prog.\ Theor.\ Phys.\ {\bf 33}, 769 (1965).
\bibitem{Rei68} R.V.\ Reid, Jr., Ann.\ Phys.\ (NY) {\bf 50}, 411 (1968);
         B.D.\ Day, Phys.\ Rev.\ C {\bf 24}, 1203 (1981).
\bibitem{Tou75} R.\ de Tourreil, B.\ Rouben, and D.W.L.\  Sprung,
         Nucl.\ Phys.\ {\bf A242}, 445 (1975);
         J.\ C\^ot\'e, B.\ Rouben, R.\ de Tourreil, and
         D.W.L.\ Sprung, {\it ibid.} {\bf A273}, 269 (1976).
\bibitem{Lac80} M.\ Lacombe, B.\ Loiseau, J.M.\ Richard, R.\ Vinh Mau,
         J.\ C\^ot\'e, P.\ Pir\`es, and R.\ de Tourreil,
         Phys.\ Rev.\ C {\bf 21}, 861 (1980).
\bibitem{Lag81} I.E.\ Lagaris and V.R.\ Pandharipande,
         Nucl.\ Phys.\ {\bf A359}, 331 (1981).
\bibitem{Sto94} V.G.J.\ Stoks, R.A.M.\ Klomp, C.P.F.\ Terheggen,
         and J.J.\ de Swart, Phys.\ Rev.\ C {\bf 49}, 2950 (1994).
\bibitem{Nag78} M.M.\ Nagels, T.A.\ Rijken, and J.J. de Swart,
         Phys.\ Rev.\ D {\bf 17}, 768 (1978).
\bibitem{said} R.A.\ Arndt, Scattering Analysis Interactive Dial-in
         (SAID), Virginia Polytechnic Institute and State University,
         status of September 1994.
\bibitem{Lis82} P.W.\ Lisowski, R.E.\ Shamu, G.F.\ Auchampaugh,
         N.S.P.\ King, M.S.\ Moore, G.L.\ Morgan, and T.S.\ Singleton,
         Phys.\ Rev.\ Lett.\ {\bf 49}, 255 (1982).
\bibitem{Wir94} R.B.\ Wiringa, V.G.J.\ Stoks, and R.\ Schiavilla,
         Phys.\ Rev.\ C, in press.
\end{references}
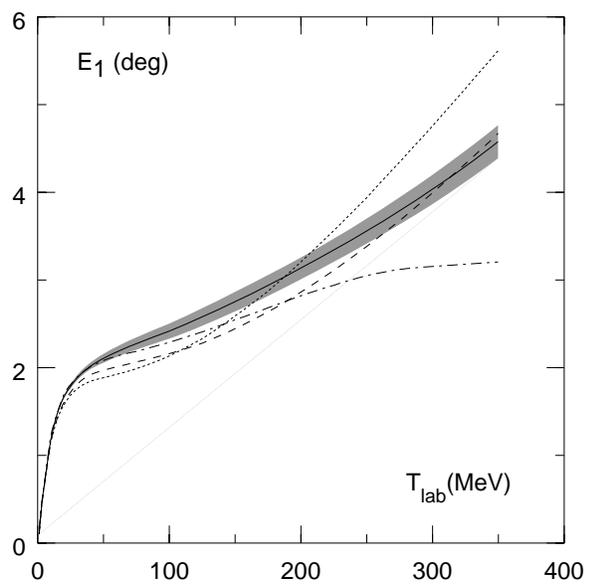
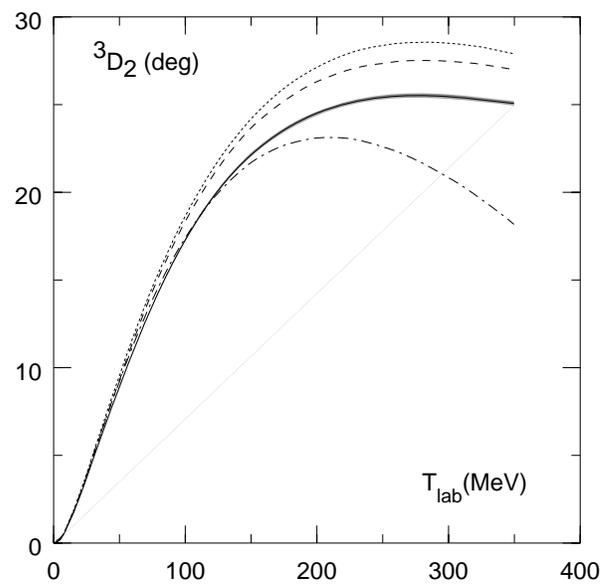
\end{document}